\documentclass{llncs}
\usepackage{times}
\usepackage[pdftex]{color,graphicx}
\usepackage{subfig}
\usepackage{amsmath,amsfonts}
\usepackage{mathpartir}

\newcommand{\true}{\ensuremath{\mathtt{true}}}
\newcommand{\false}{\ensuremath{\mathtt{false}}}
\newcommand{\goesto}[1][]{\stackrel{#1}{\longrightarrow}}

\begin{document}
%\graphicspath{{fig/}}

\title{On the Simulation of Time-Triggered Systems on a Chip with BIP}
\author{Jan Olaf Blech\inst{1}, Benoit Boyer\inst{2}, Thanh Hung Nguyen\inst{2}}
\institute{fortiss GmbH, Munich~\inst{1} \\
Verimag, Grenoble~\inst{2} }

\maketitle
\begin{abstract}
  In this report, we present functional models for software and hardware components of
  Time-Triggered Systems on a Chip (TTSoC). These are modeled in the
  asynchronous component based language BIP. We demonstrate the
  usability of our components for simulation of software which is
  developed for the TTSoC. Our software comprises services and an
  application part. Our approach allows us to simulate and validate
  aspects of the software system at an early stage in the development
  process and without the need to have the TTSoC hardware at hand.
\end{abstract}

\section{Introduction}

Simulation and analysis of systems at an early stage in the
development process allows the identification of problems prior to the
systems deployment. Thus, it can save development costs.

In this paper we present an approach that allows the simulation of
application software parts of embedded systems without the hardware
and low level software drivers and operating system. In particular we
are targeting systems based on Time-Triggered Systems on a Chip
(TTSoC)~\cite{paukovits08} hardware.  TTSoC are multi core systems
where hosts -- usually comprising at least a core and local memory --
communicate at pre-defined periodic times with each other. All hosts
are integrated on one chip. Thus, it is possible to achieve time
guarantees for messages sent between different hosts. This guaranteed
behavior facilitates certification, e.g., in the automotive or
avionics industry.  Cores may be specialized, e.g., for application
code -- our deployment scenarios typically feature one piece of
software which controls the rest of the system called application --
and I/O. We formally describe an abstract model of the TTSoC in
software using the BIP (Behavior, Interaction, Priority~\cite{bip})
modeling language.  This allows us to simulate software parts of the
system prior to the deployment on the hardware. The deployment may be
an expensive process, in some cases the hardware might even not be
available at the start of a development project. Using our approach we
are able to simulate software and hardware-parts of a system in
software. This enables us to test software which can interact with the
simulated software and hardware parts (software-in-the-loop). Our BIP
model allows the simulation of application and I/O communication parts
running on different cores.  Our simulation aims particularly at causal
dependencies between components and their interactions.  These aspects
behave in the same way as in the non-simulated system.  Causal
dependencies is an important aspect in multi-core systems and even
more crucial in TTSoC based systems.  Furthermore, our models represent
some architectural features of the system.

\begin{figure} %{c}{1.2\textwidth}
  \centering
  \includegraphics[width=0.55\textwidth]{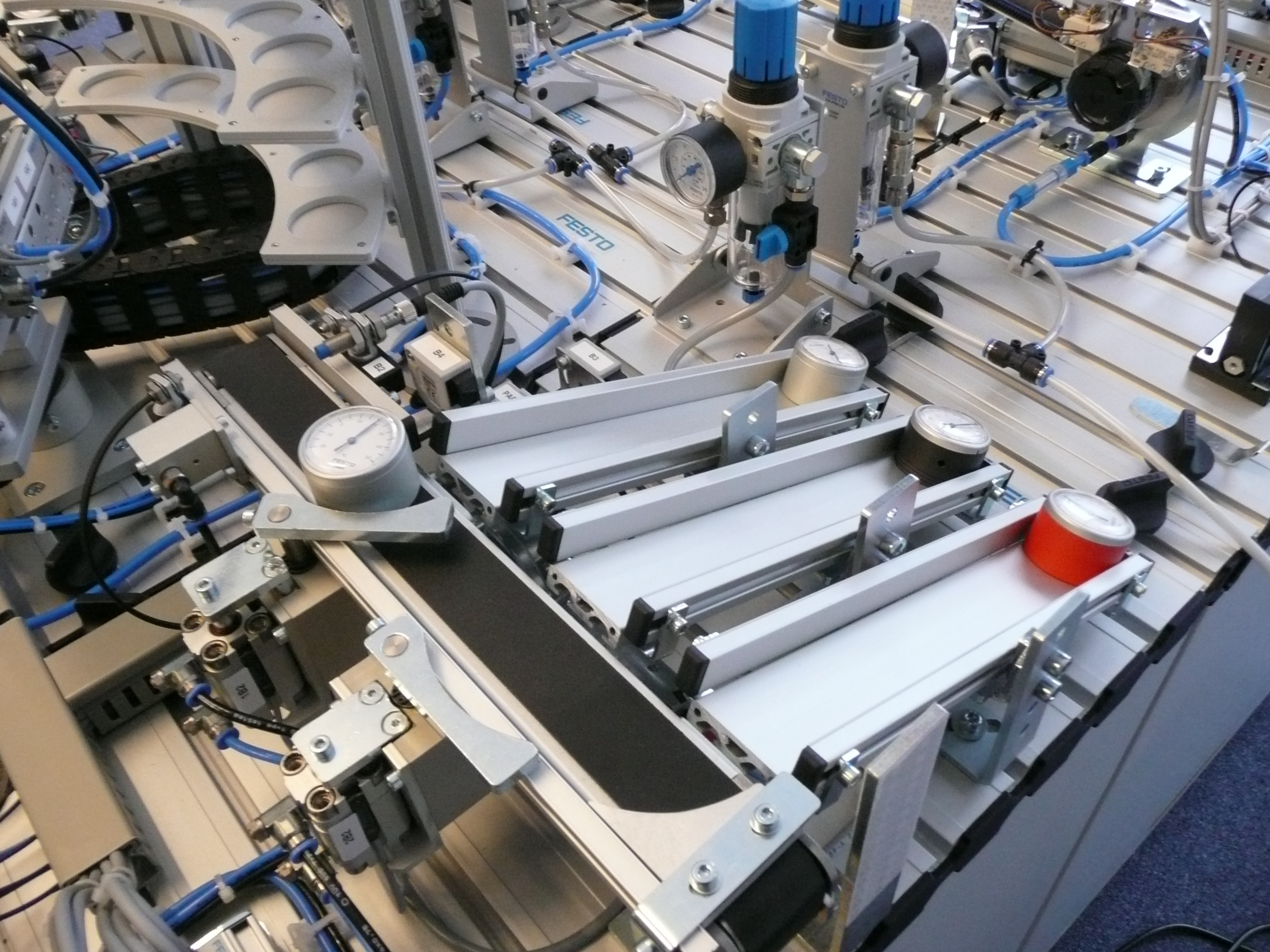}
  \caption{Sorting machine overview}
  \label{fig:casestudypc}
\end{figure}

In this paper we target TTSoC systems for controlling industrial
automation devices. As a case study we are describing the BIP based
simulation code which is needed to simulate the application software
which controls a sorting station (Figure~\ref{fig:casestudypc}) used
in the industrial automation domain.
The main contributions of this paper are the BIP models of the TTSoC
system, the case study, and a method to integrate generated C code
pieces from tool chains used in the industrial automation domain into
BIP models that simulate the industrial automation domain.

\subsection{Related Work}

The Distribution Operation Layer (DOL)~\cite{{dol}} is used for the
analysis of embedded multiprocessor systems. DOL can be used for
system performance analysis as well as optimization / design space
exploration tasks, like scheduling of applications. The Unified
Modeling Language (UML)~\cite{uml2}, a system modeling language, is
used to specify, construct, modify and visualize object-oriented
software systems. Another approach for modeling and simulating
real-time embedded systems is developed in the Ptolemy project
\cite{Ptolemy}.  Furthermore, a language originally driven by the
avionics industry for simulating for the analysis and specification of
hardware and software architectures is AADL~\cite{aadl}.  For other
languages and notations specific to tools, we can mention Simulink /
Stateflow that is used to model and simulate event-driven systems;
SystemC~\cite{systemc}, a standard design and verification language
built in C++; Metropolis~\cite{metropolis}, an environment for complex
heterogeneous electronic system design that supports simulation,
verification and synthesis; and IF-toolset~\cite{if}, an environment
for modeling and validation of heterogeneous real-time systems.

In contrast to our work carried out using BIP the real-time aspects
and precise timing conditions are of greater importance in these
approaches. Thus, our models are more abstract and simpler to
use. Simulating systems on the more abstract level is justified by the
fact that we do not know the full timing properties of our system at
simulation time. Hardware specifications might also be subject to
change at the time we run our simulations. In our work we rather want
to find out possible constraints that need to be fulfilled by running
a randomized simulation. These constraints are taken into account
during the implementation by, e.g., ensuring that certain code parts
meet an upper bound execution deadline by using a Worst-Case Execution
Time Analysis tool.

A formal study and modeling of some aspects of the same sorting
station from the industrial automation domain that we describe in this
paper can be found in~\cite{blechsidi}. The Coq theorem
prover~\cite{coq} is used to prove some properties of the IEC 61131--3
model~\cite{iec61131} of this same station. However, TTSoC aspects are
not regarded in this Coq based work.

\subsection{Overview}
An overview on the BIP language is given in
Section~\ref{sec:bip}. Section~\ref{sec:ttsoc} describes
Time-Triggered Systems on a Chip and Section~\ref{sec:modelttnoc}
presents their modeling in BIP. A case study simulating the
application software controlling an industrial automation device is
featured in Section~\ref{sec:casestudy}. A short evaluation is given
in Section~\ref{sec:eval}. A conclusion is given in
Section~\ref{sec:conclusion}.

\section{BIP - Behavior Interaction Priority}
\label{sec:bip}
%%%%%%%%%%%%%%%%%%%%%%%%%%%%%%%%%%%%%%%%%%%%%%%%%%%%%%%%%%%%%%%%%%%%%%%%%%%%%%%%%%%%%%%%%%%%%%%
%%%%%%%%%%%%%%%%%%%%%%%%%%%%%%%%%%%%%%%%%%%%%%%%%%%%%%%%%%%%%%%%%%%%%%%%%%%%%%%%%%%%%%%%%%%%%%%
%
%\vspace*{-0.5em}
%
\newcommand{\D}{\mathbb{D}}
\newcommand{\B}{\mathbb{B}}
In this section we recall the necessary concepts of the BIP framework
\cite{bip}. BIP is a component-based framework for constructing
systems by superposing three layers of modeling: Behavior,
Interaction, and Priority. The \emph{behavior} layer consists of a set
of atomic components represented by transition systems. The
\emph{interaction} layer models the collaboration between
components. Interactions are described using sets of ports and
connectors between them. The \emph{priority} layer is used to enforce
scheduling policies applied to the interaction layer, given by a
strict partial order on interactions.
%
% \vspace{-1em}
%%%%%%%%%%%%%%%%%%%%%%%%%%%%%%%%%%%%%%%%%%%%%%%%%%%%%%%%%%%%%%%%%%%%%%%%
%%%%%%%%%%%%%%%%%%%%%%%%%%%%%%%%%%%%%%%%%%%%%%%%%%%%%%%%%%%%%%%%%%%%%%%%
\subsection{Component-based Construction}
%%%%%%%%%%%%%%%%%%%%%%%%%%%%%%%%%%%%%%%%%%%%%%%%%%%%%%%%%%%%%%%%%%%%%%%%
%%%%%%%%%%%%%%%%%%%%%%%%%%%%%%%%%%%%%%%%%%%%%%%%%%%%%%%%%%%%%%%%%%%%%%%%
%
BIP offers primitives and constructs for modeling and composing
complex behaviors from atomic components. Atomic components are
Labeled Transition Systems (LTS) extended with C/C++ functions and
data. Transitions are labeled with sets of communication
ports. Composite components are obtained from atomic components by
specifying connectors and priorities.
\vspace{-1em}
\subsubsection{Atomic Components}
An atomic component is endowed with a set of local variables $X$
taking values in a domain $\D$. A valuation of the set $X$ is a function
of $X \rightarrow \D$ that maps each variable to a value. Atomic components
synchronize and exchange data with other components through the notion
of \emph{port}.
\begin{definition}[Port]
  A port $p[X']$, where $X'\subseteq X$, is defined by a port
  identifier $p$ and some data variables in a set $X'$ (referred as
  the support set).
\end{definition}
\vspace*{-0.5em}
\begin{definition}[Atomic component]
  An atomic component $B$ is defined as a tuple $(P,L,$ $T,X,\{g_\tau\}_{\tau \in T}, \{f_{ \tau}\}_{\tau \in T})$,
  where: %\vspace*{-0.25cm}
  \begin{itemize}
  \item $(P,L,T)$ is an LTS over a set of ports $P$. $L$ is a set of
    control locations and $T \subseteq L \times P \times L$ is a set of
    transitions.
  \item $X$ is a set of variables.
  \item For each transition $\tau \in T$:
    \begin{itemize}
    \item $g_\tau $ is a boolean condition over a valuation of $X$: the guard of $\tau$,
    \item $f_\tau$ is the computation step of $\tau$, a list of statements.
    \end{itemize}
  \end{itemize}
\end{definition}
\vspace{-0.5em}
For $\tau = (l,p,l')\in T$ a transition of the
internal LTS, $l$ (resp. $l'$) is referred as the source (resp.
destination) location and $p$ is a port through which an interaction
with another component can take place. Moreover, a transition $\tau =
(l,p,l')\in T$ in the internal LTS involves a transition in the atomic
component of the form $(l,p,g_\tau,f_\tau,l')$ which can be executed
only if the guard $g_\tau$ evaluates to $\true$, and $f_\tau$ is a
computation step consisting of transformations of local variables in $X$.
\vspace{-0.5em}
\paragraph{Example: Atomic component ``Global timer''}

\begin{figure}
\begin{center}
\resizebox{3cm}{!}{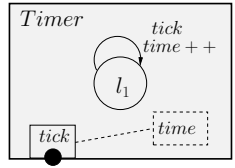}
\caption{\footnotesize Atomic component}
\label{fig:atom}
\end{center}
\end{figure}

Figure~\ref{fig:atom} shows the global timer (clock) used in our TTSoC that we modeled as an example of an atomic component.

This atomic component has a port $tick$, a control location $l_1$ and
a variable $time$ that is associated to the port $tick$ and is
increased every time the transition tick occurs. The absence of guard
on the transition $tick$ implies that its guard is always true.

\paragraph{Semantics of Atomic Components.}
The semantics of an atomic component is an LTS over configurations and ports, formally defined as follows:
\begin{definition}[Semantics of Atomic Components]
The semantics of the atomic component $(P,L,T, X, \{g_{\tau}\}_{\tau \in T}, \{f_{\tau}\}_{\tau \in T})$ is an LTS $(P,Q,T')$ s.t. \vspace*{-0.25cm}
\begin{itemize}
\item $Q = L\times [X\rightarrow \D]$,
\item $T'= \{ ((l,v),p, (l',v'))\in Q\times P\times Q\mid \exists \tau = (l, p, l') \in T: g_\tau(v) \land v' = f_\tau(v)\}$.
\end{itemize}
\end{definition}
A configuration is a pair $(l,v)\in Q$ where $l \in L$ is a control
location, and $v \in [X \rightarrow \D]$ is a valuation of the
variables in $X$. The evolution of configurations $(l_1, v)
\xrightarrow{p(v_p)} (l_2, v')$, where $v_p$ is a valuation of
variables attached to port $p$, is possible if there exists a
transition $(l_1, p[X_p], g_\tau, f_\tau, l_2)$,
s.t. $g_\tau(v)=\true$. As a result, the valuation $v$ of variables is
modified to $v'=f_\tau(v[X_p \leftarrow v_p])$.
\vspace{-1em}
\subsubsection{Creating composite components.} Assuming some available
atomic components $B_1,$ $\ldots,B_n$, we show how to connect
$\{B_i\}_{i\in I}$ with $I\subseteq [1,n]$ using \emph{connectors}.

A connector $\gamma$ is used to specify possible interactions,
i.e. the sets of ports that have to be jointly executed. Two types of
ports (\textit{synchron}, \textit{trigger}) are defined, in order to
specify the feasible interactions of a connector. A \textit{trigger}
port is active: it can initiate an interaction without synchronizing
with other ports. It is represented graphically by a triangle. A
\textit{synchron} port is passive: it needs synchronization with other
ports for initiating an interaction. It is denoted by a circle. A
feasible interaction of a connector is a subset of its ports
s.t. either it contains some trigger, or it is maximal.

\begin{figure}
  %\hfill
\begin{center}
  \begin{minipage}{.95\textwidth}
    % \begin{center}
\begin{center}
    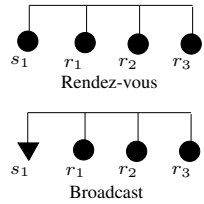
\end{center}
    \caption{ Rendez-vous and broadcast}
    \label{fig:rendezvous-broadcast}
    % \end{center}
  \end{minipage}
\end{center}
  %\hfill
\end{figure}

\begin{figure}
  \begin{minipage}{0.95\textwidth}
    % \begin{center}
    \includegraphics[width=1.0\textwidth]{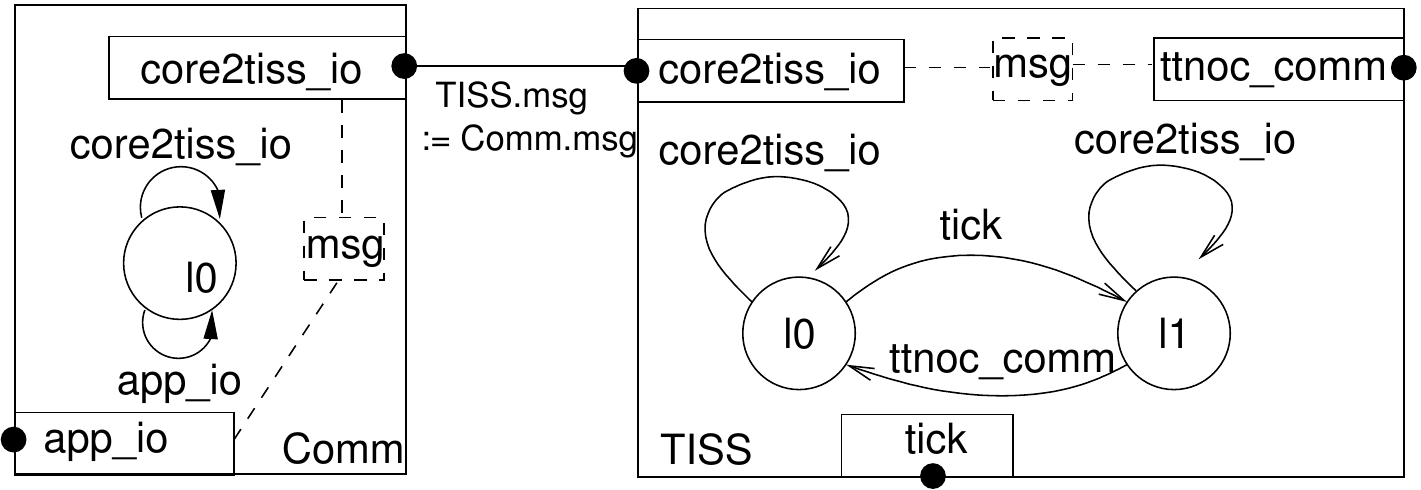}
    \caption{ A TTSoC connector with data transfer}
    \label{fig:ttsoc-conn}
    % \end{center}
  \end{minipage}
  \hfill
\end{figure}

Figure~\ref{fig:rendezvous-broadcast} shows two connectors:
\textit{Rendezvous} (only the maximal interaction $s_1r_1r_2r_3$ is
possible), \textit{Broadcast} (all the interactions containing the
trigger port $s_1$ are possible).

\noindent
Formally, a connector is defined as follows:
\begin{definition}[Connector]
  \label{def:connector}
  A connector $\gamma$ is a tuple $({\cal P}_\gamma, t, G,F)$,
  where:\vspace*{-0.25cm}
  \begin{itemize}
  \item ${\cal P}_\gamma = \{p_i[X_i]\mid p_i\in B_i.P\}_{i \in I}$
    s.t. $\forall i\in I: {\cal P}_\gamma \cap B_i.P = \{p_i\}$,
  \item $t: {\cal P}_\gamma\rightarrow \B$ s.t. $t(p)=\true$ if $p$ is
    trigger (and $\false$ if synchron),
  \item $G$ is a Boolean function over the set of variables
    $\cup_{i\in I} X_i$ (the guard),
  \item $F$ is an update function defined over the set of variables
    $\cup_{i\in I} X_i$.
  \end{itemize}
\end{definition}
${\cal P}_\gamma$ is the set of connected ports called the support set
of $\gamma$. The ports in ${\cal P}_\gamma$ are tagged with function
$t$ indicating whether they are trigger or synchron. Moreover, for
each $i\in I$, $x_i$ is a set of variables associated to the port
$p_i$.

A communication between the atomic components of $\{B_i\}_{i\in I}$
through a connector $({\cal P}_\gamma, G,F)$ is defined using the
notion of \emph{interaction}:
\begin{definition}[Interaction]
\label{def:int}
A set of ports $a=\{p_j\}_{j\in J} \subseteq {\cal P}_\gamma$ for some
$J\subseteq I$ is an interaction of $\gamma$ if one of the following
conditions holds: (1) there exists $j \in J$ s.t. $p_j$ is trigger;
(2) for all $j \in J$, $p_j$ is synchron and $\{p_j\}_{j\in J} = {\cal
  P}_\gamma$.
\end{definition}
An interaction $a$ has a guard and two functions $G_a,F_a$,
respectively obtained by projecting $G$ and $F$ on the variables of
the ports involved in $a$. We denote by ${\cal I}(\gamma)$ the set of
interactions of $\gamma$. Synchronization through an interaction
involves two steps. First, the guard $G_a$ is evaluated, then the
update function $F_a$ is applied. If there are several possible
interactions inside a connector, we choose the interaction involving
the maximum\footnote{If there are several maximal interactions, the
  choice between them is at random.} number of ports. One can also add
priorities to reduce non-determinism whenever several interactions are
enabled. Then, the interaction with the highest priority is chosen.

In the TTSoC system that we modeled, the global timer communicates
with all the components that need to synchronize their action
according to some time schedule. These communications are done by
using interactions between the global timer and these
components. Figure~\ref{fig:ttsoc-conn} represents a connector with
data transfer used in the TTSoC model. It connects two ports
$core2tiss\_io$ of a communication service component $Comm$ and of a
TISS component. These ports have their own associated message
variables $msg$. The message variable of the $Comm$ component is sent
over the connector to the $TISS$ component ($TISS.msg = Comm.msg$).

\begin{definition}[Composite Component]
  A composite component is defined from a set of available atomic
  components and a set of connectors. The connection of the
  $\{B_i\}_{i\in I}$ using the set $\Gamma$ of connectors is denoted
  $\Gamma(\{B_{i}\}_{i\in I})$.
\end{definition}
Note that a composite component obtained by composition of a set of
atomic components can be composed with other components in a
hierarchical and incremental fashion using the same operational
semantics.
\begin{definition}[Semantics of Composite Components]
\label{def-runtimesemanticscomposite}
A state $q$ of a composite component $C = \Gamma(B_1, \ldots, B_n)$,
where $\Gamma$ connects the $B_i$'s for $i\in I$, is a $n$-tuple
$q=(q_1,\ldots,q_n)$ where $q_i=(l_i,v_i)$ is a state of $B_i$. Thus,
the semantics of $C$ is precisely defined as a transition system
$(Q,A,\goesto)$, where: \vspace*{-0.25cm}
\begin{itemize}
\item $Q= B_1.Q\times \ldots\times B_n.Q$,
\item $A = \{a \in {\cal I}(\gamma)\}_{\gamma \in \Gamma}$ is the set of all possible interactions,
\item $\goesto$ is the least set of transitions satisfying the following rule:
\vspace*{-0.25cm}
\begin{mathpar}
\inferrule*
{
    \exists\gamma \in \Gamma: \gamma = (P_\gamma,G,F) \and \exists a \in {\cal I}(\gamma)\and
    G_a(v(X)) \hva\\
    \forall i\in I:\ q_i \goesto[p_i(v_i)]_i q'_i \wedge v_i = F_{ai}(v(X)) \and
    \forall i\not\in I:\ q_i = q'_i
}
{
    (q_1,\dots,q_n) \goesto[a] (q'_1,\dots,q'_n)
}
\end{mathpar}
where $a = \{p_i\}_{i \in I}$, $X$ is the set of attached variables on the ports
of $a$, $v$ is the global valuation of variables, and $F_{ai}$ is the
partial function derived from $F$ restricted to the variable
associated to $p_i$.
\end{itemize}
\end{definition}
The meaning of the above rule is the following: if there exists an
interaction $a$ s.t. all its ports are enabled in the current state
and its guard ($G_a(v(X))$) is true, then we can fire the
interaction. When $a$ is fired, not involved components stay in the
same state, and, involved components evolve according to the
interaction.

Notice that several distinct interactions can be enabled at the same
time, thus introducing non-determinism in the product behavior,
possibly restricted using priorities.
\begin{definition}[Priority]
  \label{defn:priority}
  Let $C = (Q,A,\goesto)$ be the behavior of the composite component
  $\Gamma(B_1, \ldots, B_n)$.  A {\em priority model} $\pi$ is a
  strict partial order on the set of interactions $A$. Given a
  priority model $\pi$, we abbreviate $(a,a')\in \pi$ to $a \prec
  a'$. The component $\pi(C)$ is defined by the behavior $(Q, A,
  \goesto_\pi)$, where $\goesto_\pi$ is the least set of transitions
  satisfying the following rule: \vspace*{-0.25cm}
\begin{mathpar}
\inferrule*{
  q \goesto[a] q' \and
  \not\exists a'\in A,\exists q''\in Q: a \prec a' \land q \goesto[a'] q''
}{
  q \goesto[a]_\pi q'
}
\end{mathpar}
\end{definition}
An interaction is enabled in $\pi(C)$ only if it is enabled in $C$, and,
it is maximal according to $\pi$ among the active interactions in $C$.

Finally, we consider systems defined as a parallel composition of components together with an initial state.
\begin{definition}[System]
A system ${\cal S}$ is a pair $(B,\mathit{Init})$ where $B$ is a component and $Init$ is the initial state of $B$.
\end{definition}

\section{Time-Triggered Systems on a Chip}
\label{sec:ttsoc}
In a TTSoC several hosts communicate with each other using a
time-triggered network. Hosts and time-triggered network are
integrated on one chip.

In this paper, we follow the TTSoC description given
in~\cite{paukovits08}. A TTNoC consists of the following components:
\begin{itemize}

\item {\bf Hosts} are physical entities that interact via a
  time-triggered network with each other. In many cases a host is a
  CPU {\bf Core} equipped with its own memory and possible local I/O
  access.  { Cores} provide computation power. Distinct cores can be
  used for handling different I/O tasks. Apart from cores, hosts can
  be connections to other bus systems or related I/O devices.
\item Hosts are connected via a Time-Triggered Network on a Chip ({\bf
    TTNoC}). \\ The TTNoC provides communication channels between the
  hosts. For each application purpose the TTNoC is configured in a way
  such that messages of fixed length can be sent between the different
  cores in distinct time slots. In our case
  (following~\cite{paukovits08}) we are looking at a TTNoC which is
  organized using a mesh structure. This means that different parts of
  communication channels are connected via switches which route
  messages through the network. One consequence -- and advantage over bus like structures -- is that unlike in
  traditional bus-systems different hosts may be communicating at the
  same time as long as their communication channel parts and switches
  do not interfere with each other.
\item The connection between a host and a TTNoC is guarded using a
  Trusted Interface Subsystem ({\bf TISS}) which serves as an
  interface and intermediate storage for the host thereby abstracting
  some TTNoC details and ensuring that time slots and routes for
  messages are met.
\end{itemize}

\begin{figure}
  \centering
  \includegraphics[width=0.95\textwidth]{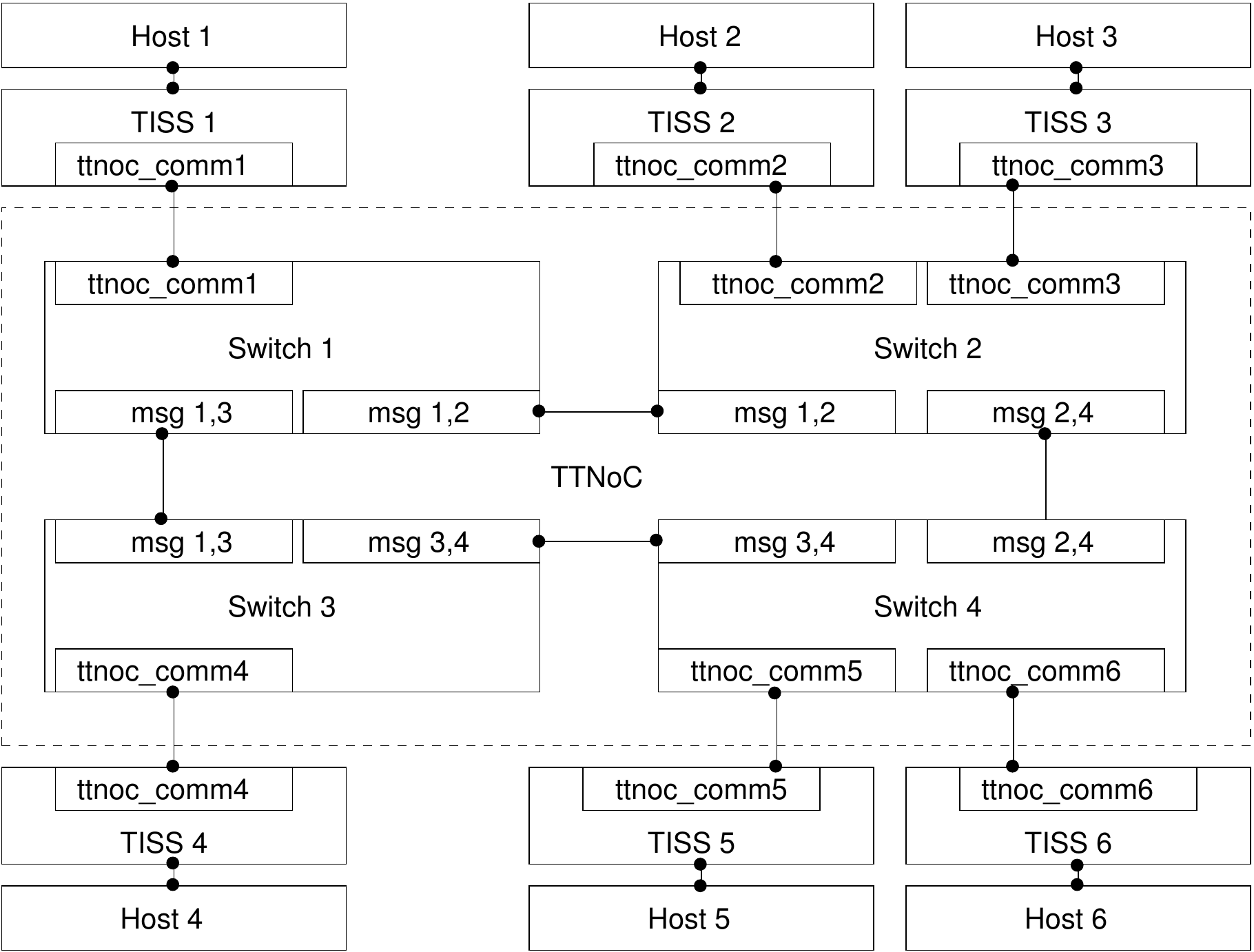}
  \caption{TTSoC  overview}
  \label{fig:ttnocbipnew}
\end{figure}

An example TTSoC is shown in Figure~\ref{fig:ttnocbipnew}. One can see
that six hosts are connected via TISS to the TTNoC. Two of the
switches are directly connected to two TISS. The other two switches
are each connected to only one TISS. The switches are connected with
each other realizing a 2x2 grid. One can see that parallel
communication is in some cases possible, e.g., Host 1 with Host 4,
Host 5 with Host 6 and Host 2 with Host 3 can exchange messages in the
same time slot.

\section{Modeling the TTNoC in BIP}
\label{sec:modelttnoc}
Here we give a description of the TTNoC components and their connection to the environment using TISS. We present their modeling using BIP.

\subsection{Managing Time}
Time-triggered systems are characterized by the fact that the
communication between different hosts is done in a synchronous
time-triggered way whereas the hosts themselves may internally behave
in an asynchronous way and the interaction with the TISS may also
behave asynchronously.
\paragraph{Global Time} Our model features a component which emits a global time tick
(cf. Figure~\ref{fig:atom}). Different parts of our BIP model can use
this time tick, e.g., for synchronization.
\paragraph{Splitting of the global time tick into subticks} In the
TTNoC BIP model at hand a global time tick {\sf tick} is followed by
three subticks {\sf t1}, {\sf t2} and {\sf t3} that represent internal
steps that are taken to transmit a message between different TISS via
the TTNoC switches. Thus, we have four ticks which may be used to
transmit a message between a TISS and a switch, transmit a message
between this switch and another switch, transmit a message between
this other switch and yet another switch, and finally transmit it to
another TISS. Thus, routes through the TTNoC may comprise at most
three switches. The time tick splitting is modeled as an independent
BIP component. Larger TTNoC would require the modeling of additional
subticks.

\subsection{The TISS}
\begin{figure}
  \centering
  \includegraphics[width=0.95\textwidth]{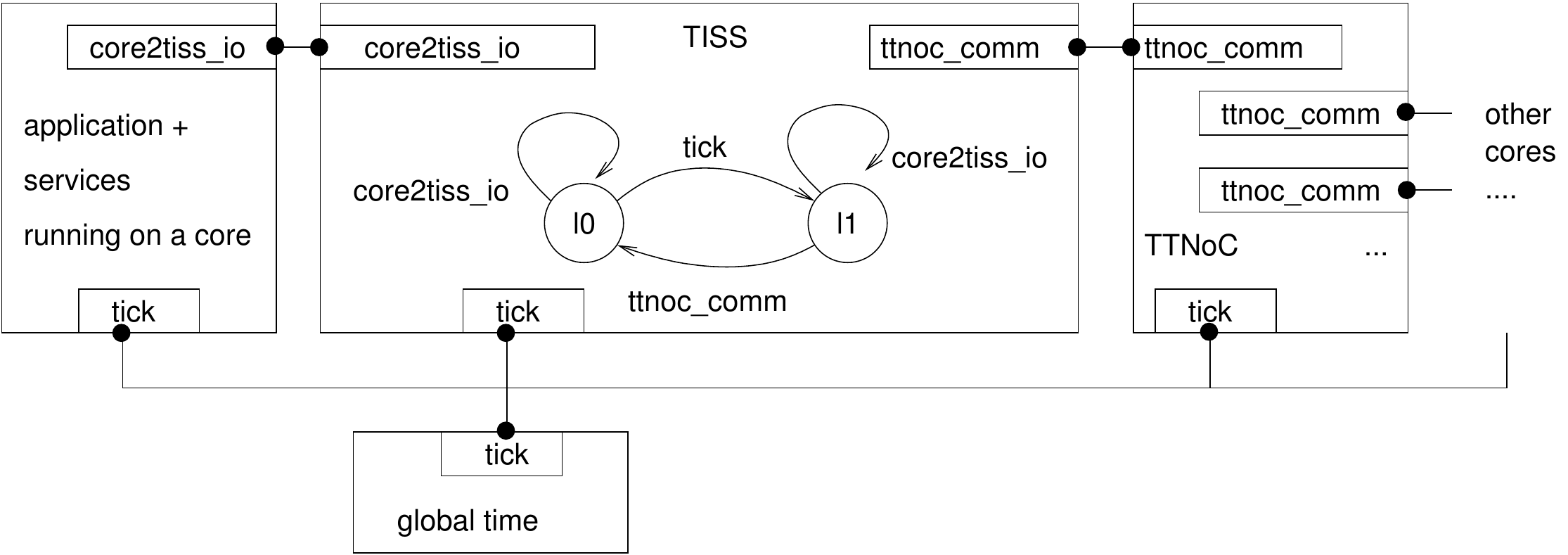}
  \caption{Connecting a host system to the TTNoC via a TISS}
  \label{fig:bip1}
\end{figure}

A TISS has two main purposes:
\begin{itemize}
\item It communicates with the host and serves as an intermediate
  storage for messages. The interface to the host associates messages
  with a port number. The interface to the TISS comprises the message
  together with routing and target host information.
\item It sends and receives messages at predefined periodical points
  in time over the TTNoC. Thereby it ensures that no collision of
  messages from different TISS occur inside the TTNoC. For this reason
  a static schedule has to be computed in advance for the entire TTSoC
  and each TISS is programmed accordingly.
\end{itemize}
Figure~\ref{fig:bip1} shows a core that communicates over a TTNoC
using a TISS. Variables and their modifications are not shown.  The
TISS receives messages from the core and sends messages to the
core. In case of incoming messages from the core, routing information
is added to messages and they are transmitted over the
TTNoC. Otherwise, the routing information is deleted and the message
is given to the core. The TISS also serves as a kind of buffer, since
TTNoC and core do not have to be synchronized.
The BIP model comprises two locations {\sf l0} and {\sf l1}. Messages
may be received and collected at any time from the host system using
the connector between the {\sf core2tiss\_io} ports. The transmission
and receiving of messages to and from he TTNoC happens only at the
{\sf ttnoc\_comm} transition from {\sf l1} to {\sf l0} while the
global time tick performs a transition from {\sf l0} to {\sf l1}. This
ensures that at most one message is either sent over or received from
the TTNoC per global time tick from a single TISS. Incoming messages
from the TISS are associated with a port number and stored
intermediately. The port number is used to determine the target host
and additional routing information. This resolvement happens during or
before a {\sf ttnoc\_comm} transition. Messages that are to be collected from
the host by the TISS are also stored in the TISS together with a port number.  The
conditions when a message is actually sent during the {\sf ttnoc\_comm
} typically depends on additional internal variables that can, e.g.,
count global time ticks in the TISS during the {\sf tick}
transition. This ensure that different types of messages (e.g.,
associated with different ports) are only sent at predefined periodic
points in time to predefined targets and remain in storage otherwise.

\subsection{Inside the TTNoC}

We refer to Figure~\ref{fig:ttnocbipnew} for on overview on the BIP
components that represent a TTNoC and connectors and ports for message
passing. Not shown in this figure are the means to emit and handle
time ticks and the communication details.

\begin{figure}
  %\hfill
  \begin{minipage}[t]{.85\textwidth}
    \begin{center}
      \includegraphics[width=1\textwidth]{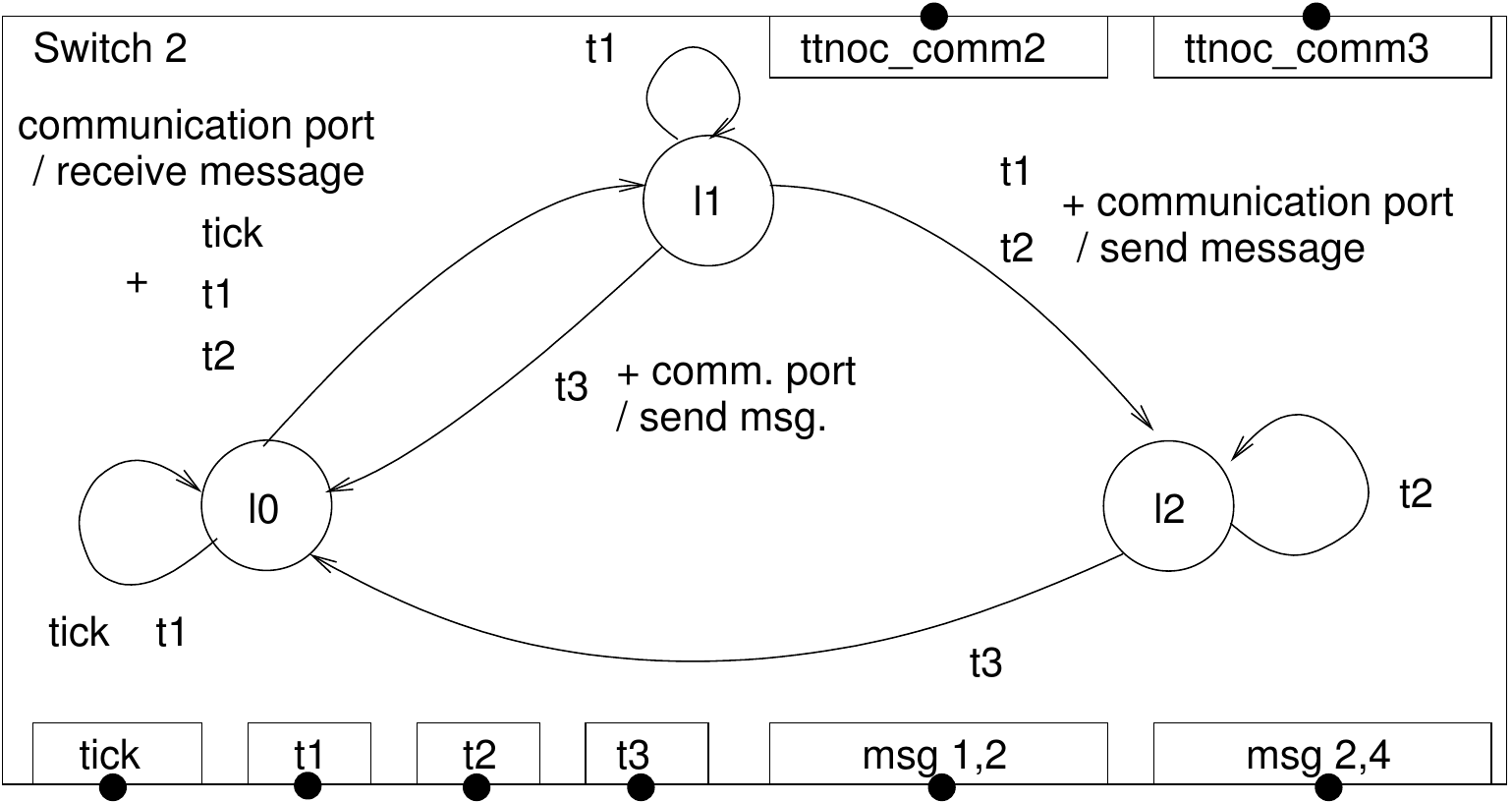}
      \caption{Inside switch 2}
      \label{fig:switch2}
    \end{center}
  \end{minipage}
  %\hfill
\end{figure}

\paragraph{BIP models for switches}
A switch can handle one message per global tick. In our BIP model for
switches in the described TTNoC we model this feature by introducing
three states: {\sf l0}, {\sf l1} and {\sf l2}. {\sf l0} is the state
before a message arrives.  {\sf l1} is the state where a message has
arrived but not transmitted, {\sf l2} represents a state where the
arrived message has been forwarded to another TISS or switch, but some
time is still remaining before the next global tick.  The arrival of a
message occurs together with {\sf tick} or during {\sf t1} or {\sf
  t2}. The routing to the other switch or TISS happens in {\sf t1},
{\sf t2} or {\sf t3}.  If it happens in {\sf t3} we return to {\sf l0}
immediately, otherwise we mark the switch as used by taking state {\sf
  l2}. Figure~\ref{fig:switch2} gives an overview on a simple BIP
model for the switch 2. It omits some communication details. The BIP
models used in a more comprehensive implementation feature additional
intermediate locations not visible to the external to facilitate the
handling of additional constraints.

\begin{figure}

  \begin{minipage}[t]{.5\textwidth}
    \begin{center}
      \includegraphics[width=1\textwidth]{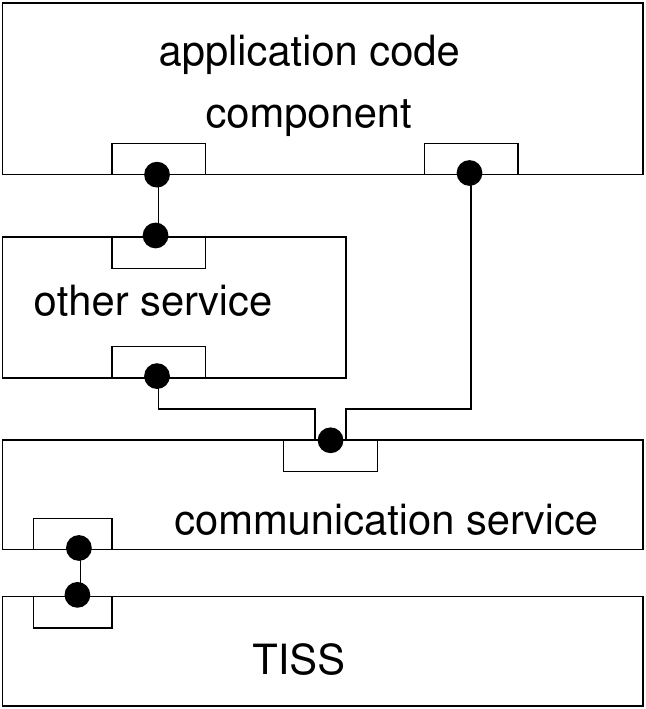}
      \caption{Application and services interactions}
      \label{fig:appcodebip}
    \end{center}
  \end{minipage}
  \hfill
\end{figure}

\paragraph{The TTSoC as a Component}
The TTSoC itself is modeled as a single composite BIP component comprising the switches, TISS components and connectors between these components. As interfaces it offers connections to the hosts and to the global time tick.

\subsection{Modeling the Host System}
In our case modeling a host system means the creation of BIP models that simulate the entire software that runs on a core and its execution characteristics.
In the proposed scenario a host system is composed of components that realize: communication services, higher-level services, application code and I/O. Thus, the host is realized using different BIP components which may interact with each other. Unlike in the TTNoC different BIP components, e.g., representing different threads on the host core can run completely asynchronous. This may, e.g., be the case for different threads running on a CPU core.

\paragraph{ Communication Services}
Communication services represent software parts that realize some functionality that the application code may use. This can comprise operating system services and special hardware drivers.  Here they are realized as BIP components that are connected to a TISS on the one side and to higher-level services, I/O, and application models on the other side. They simulate, services that we are implementing as part of a basic software support for our TTSoC.

\paragraph{Higher-level Services}
Higher-level services are  composed services that realize some higher-level functionality based on other services. Here we have modeled a voting service which takes several input values, e.g., from different sensors and establishes a mean value which is forwarded to the application.

\paragraph{Application Code Component}
The application code is modeled as an atomic or composite single BIP component.
A scenario with an application code component with connectors to other services is shown in Figure~\ref{fig:appcodebip}. %The figure omits concrete port names.
%\begin{figure}

%\end{figure}

\paragraph{Input / Output Components}
We provide BIP models that simulate Input and Output operations. These comprise simulation components that provide simulation of sensor data. Furthermore output components that simulate, e.g., actuators. In the current implementation these output components write their status data to files.

\paragraph{Realizing a Host System}

\begin{figure}
  \begin{center}
    \includegraphics[width=0.95\textwidth]{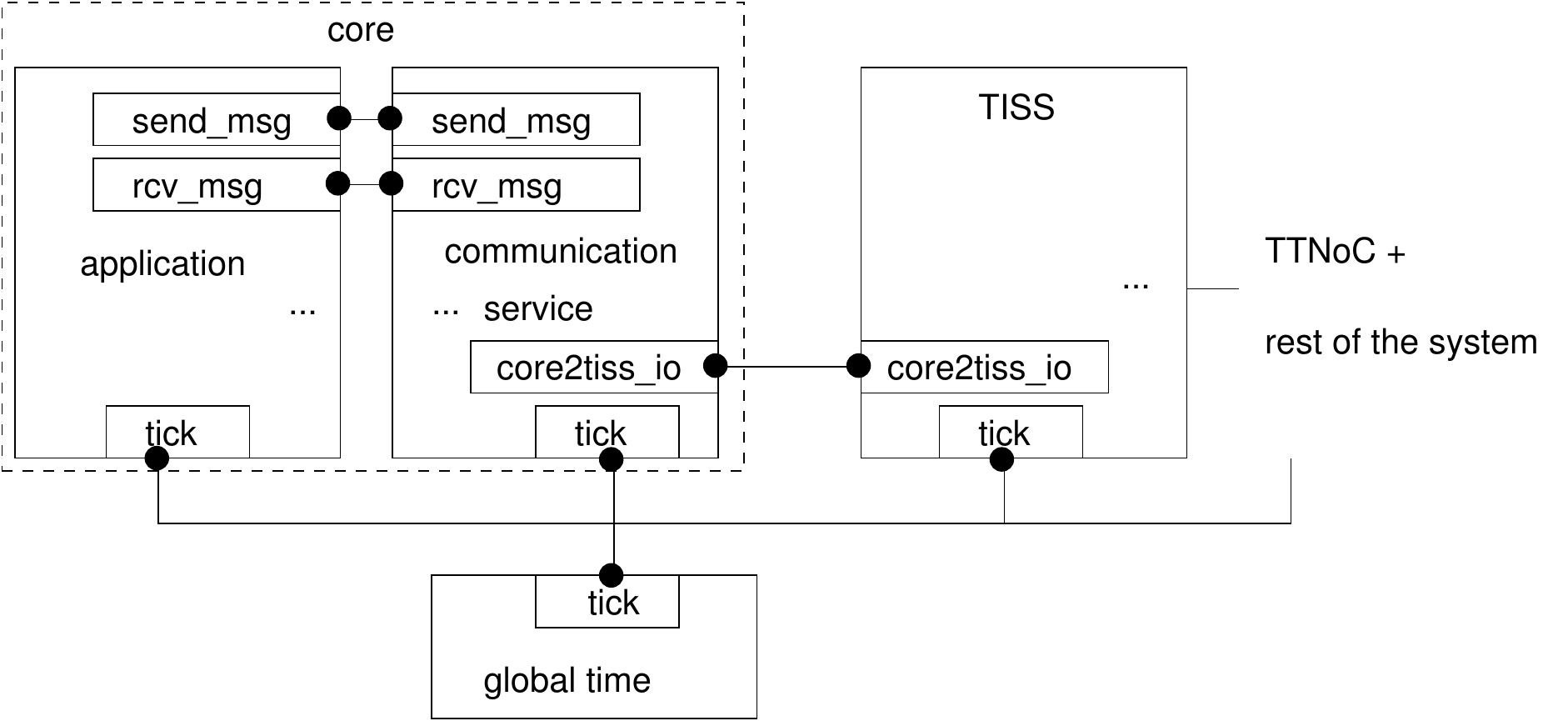}
  \end{center}
  \caption{ Application and services on a core connected to a TISS}
  \label{fig:bip2}
\end{figure}

Figure~\ref{fig:bip2} shows the interaction of an application with a communication service. This interaction may occur asynchronously to the global time tick and the TTNoC. By means of this asynchronity we model the much faster execution clock speed of a core compared to the TTNoC.

\subsection{Application Code for Industrial Automation}
 In this paper we target different application scenarios for the usage of TTSoC based systems in the industrial automation domain. Software for industrial automation is typically described using the IEC 61131--3~\cite{iec61131} standard. These software descriptions divide computation into different steps which are often executed one after the other forming a kind of loop structure with branches. C code is typically generated for each of these steps individually. In the real system it is integrated into one large C loop structure. Here we realize this loop like top level control structure in BIP as an LTS and integrate generated or hand written C code that realizes the functionality represented by the steps into the LTS transitions. This way, we have a device to test and simulate these generated or hand-written C code pieces and adapt them.

\paragraph{API calls}
The generated or hand-written C code contains API calls (e.g., a POSIX API).
Communication with BIP modeled services is done by using the same C API calls  from the application code. However, the implementation is done in a slightly different way. Our C functions store and retrieve values from intermediate stores which are filled and collected by the services. The same principle is used in the real implementation so our simulation sticks close to it. In the real implementation services can, e.g., be realized as independent threads.

\paragraph{Time}
The execution of a step is usually associated with a maximal execution time called time slice. In our model execution of the C code associated with a step is done during one state transition. Modeling the duration of this execution is done by requiring that a  number of global time ticks (corresponding to the time-slice) have to be elapsed before the application code component is able to communicate its new result and control is passed to the BIP transition associated with the next step.

We have described an additional transformation from IEC 61131--3 to BIP in~\cite{la}. Unlike in this work, here we keep the BIP structure as minimal as possible in order to simulate the C-code pieces in the most realistic way.

\section{Application and Service Simulation for a Sorting Station}
\label{sec:casestudy}
Here we describe the application code for our sorting station as depicted in Figure~\ref{sec:casestudy}.

\begin{figure}
  \begin{center}
    \includegraphics[width=0.75\textwidth]{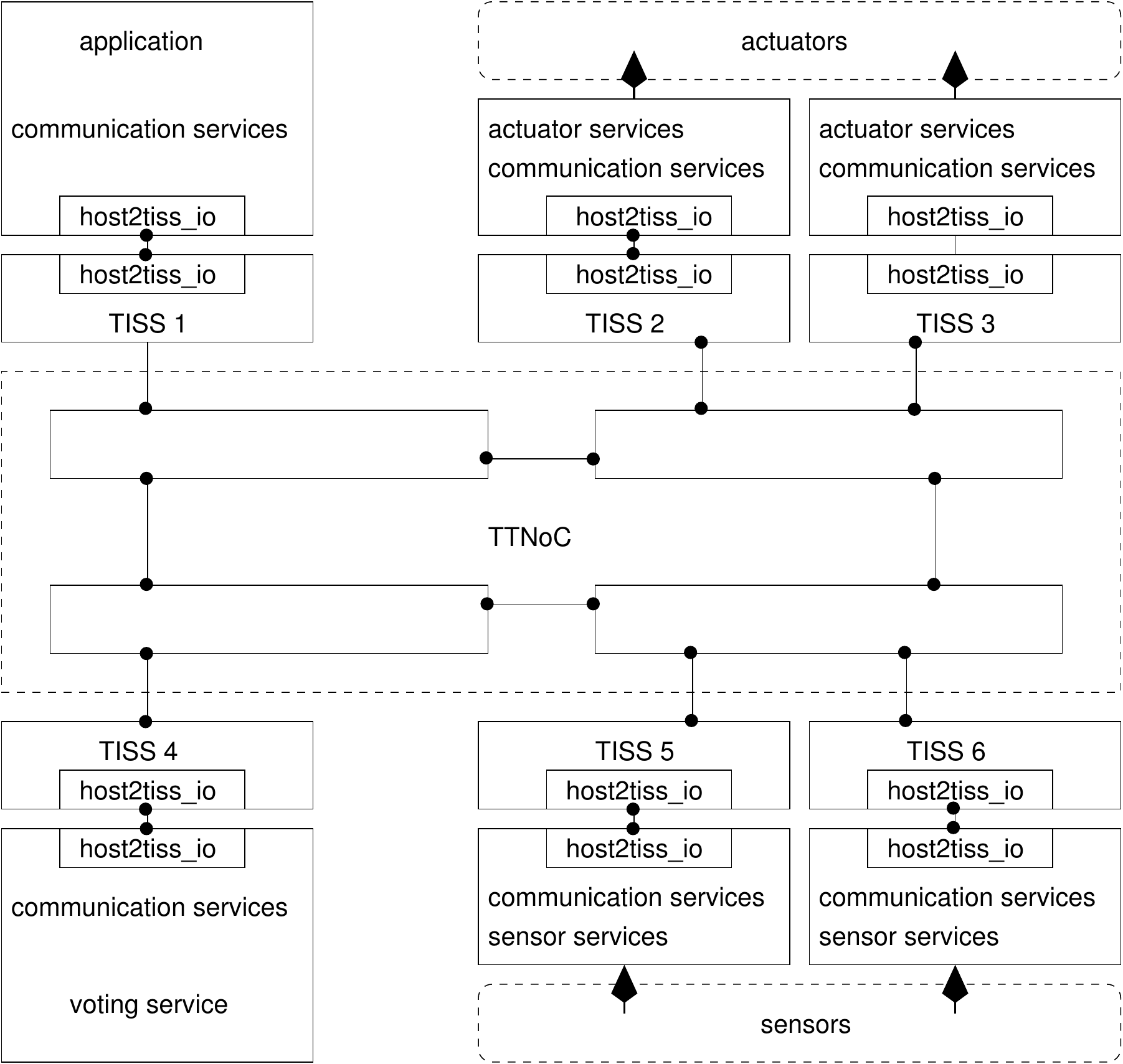}
  \end{center}
  \caption{Overview on a configuration scenario}
  \label{fig:casestudy}
\end{figure}

\begin{figure}
  \begin{center}
    \includegraphics[width=0.7\textwidth]{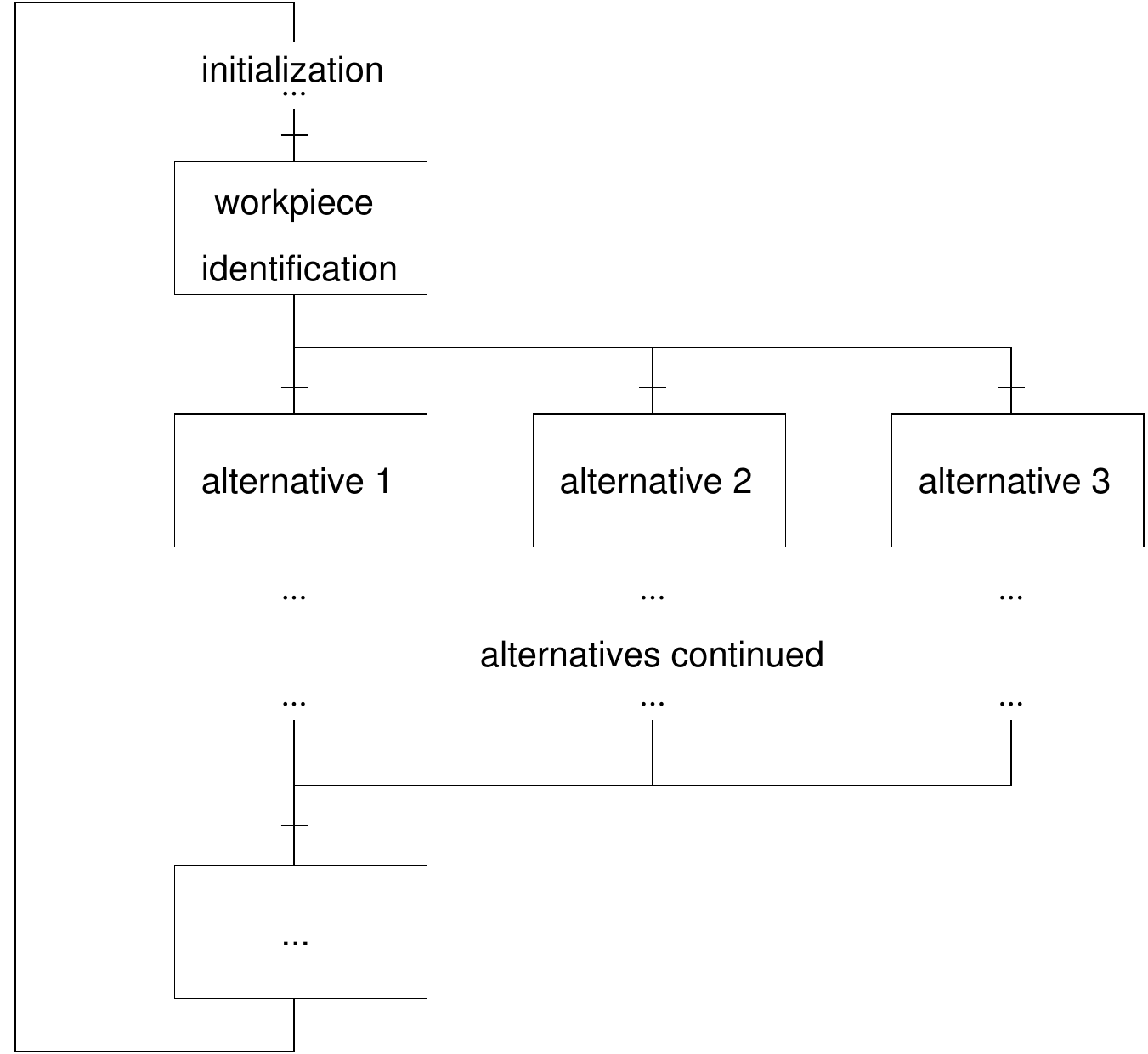}
  \end{center}
  \caption{IEC 61131--3 control structure}
  \label{fig:casestudyplc}
\end{figure}

Figure~\ref{fig:casestudy} gives an overview on a possible setting: this study  is inspired by a real existing demonstrator~\cite{D6.6}. Six hosts are connected via a TISS to the TTNoC. One host comprises a core that executes the application software, another core is dedicated to a voting service that judges the quality of values delivered by sensors.
Two hosts each perform the reading of sensor values and control of actuators. Each host features communication services to communicate via its TISS over the TTNoC with other hosts. The IEC 61131-3 structure that runs as application is sketched in Figure~\ref{fig:casestudyplc} (cf.~\cite{blechsidi}). The different steps are shown for which we integrate C-code in our BIP model of the application. The BIP model itself has a similar structure. The entire application is modeled as a single BIP component communicating with services. Transitions between steps are replaced by transitions between BIP locations. Additional transitions are inserted to handle I/O at the end of each step. Each step is modeled in a way such that it terminates in a fixed amount of time. This is a typical feature in IEC 61131--3 that we took care of here.

Common to all scenarios is that the application is running on different cores than I/O operations and has to communicate with and control the I/O.
In our real systems  the TTSoC is realized using FPGA technology. In principle it is possible to adapt the number of cores and the layout of the TTNoC to application needs. Furthermore, cores can be specialized towards distinct domains of computation and I/O.
The number of API calls in the application C-code is very small: just Posix calls to receive and transmit messages.

\section{Evaluation}
\label{sec:eval}
We have run and analyzed different system configurations using our BIP models of TTSoC components, services and application code. All of them realize different simulation scenarios for the industrial automation domain (cf. Section~\ref{sec:casestudy}). In particular they are based on our application for the sorting station shown in Figure~\ref{fig:casestudypc}.

The simulation gives us the ability to test and improve our application software. By omitting and modifying priorities of interactions different non-deterministic scenarios can be simulated. Due to this testing we where able to fix some minor errors in the actual C-code implementation.
More importantly our simulation revealed the following weakness of our overall sorting service control strategy:  it can occur that the application receives old sensor data and actuator commands are not delivered on time.
The main reason for this is that  the communication between the application and the TISS happens asynchronously and without a timing guarantee.

One solution to overcome this drawback would be to change the design of the system and establish a synchronous communication between application software and TISS at distinct points in time. This, however would require major changes in the system design.
For this reason we analyze application software parts to estimate a worst case execution time. This can be used to determine a maximal latency for reaction of the application software to sensor data and control of actuators. The overall speed of processing elements in the sorting station will be set such that these latencies do not lead to a wrong handling of an element.

\section{Conclusion}
\label{sec:conclusion}

We showed a way to simulate and validate aspects of TTSoC based systems at an early development stage.
We presented BIP models for representing hardware components of TTSoC based systems. Furthermore, we introduced BIP models for connected software services. These models provide an environment for simulating TTSoC based systems prior to deployment and availability of exact specifications. They can be used for a variety of TTSoC usage scenarios. We exemplified a case study from the industrial automation domain. Here the main purpose is simulating the controlling software parts (application) prior to availability of the entire system. Thereby we introduced a way of modeling and simulating PLC applications using BIP. Running our simulations we discovered additional timing constraints which have to be ensured in the real-implementation of the system.

As future work, we plan to investigate additional case studies in other domains. Furthermore, we are also interested to formally analyze properties of our models. These comprise analysis of invariants and related properties like deadlock freedom by using, e.g., D-Finder 2~\cite{dfinder2}. An extension to real-time aspects is also a goal. Another, area for future work is the connection of the input and output components to software that graphically displays the status of an industrial automation device, so that one can actually see a virtual video of a machine that sorts work pieces controlled by our BIP components.
\paragraph*{Acknowledgments}
This work has been supported by the European research project ACROSS under the Grant Agreement ARTEMIS-2009- 1-100208.

\bibliographystyle{alpha}

\newpage
\section*{Appendix}
\begin{figure}
  \centering
  \includegraphics[width=0.95\textwidth]{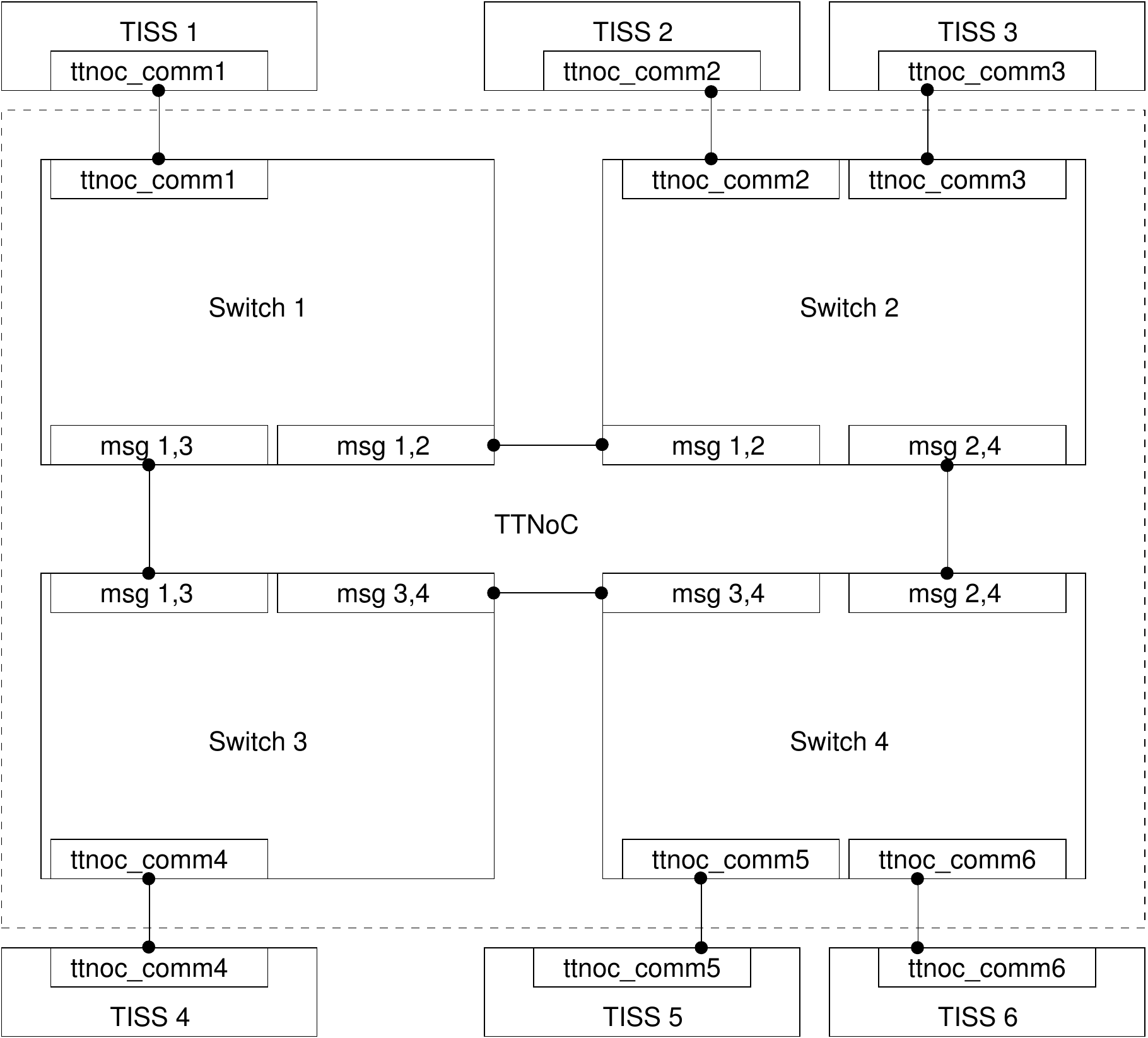}
  \caption{TTNoC model overview}
  \label{fig:a-ttnocbip}
\end{figure}

A version of our TTNoC model with port names is shown in Figure~\ref{fig:a-ttnocbip}.

%\begin{figure}
%  \centering
%  \includegraphics[width=0.95\textwidth]{TTSoC.pdf}
%  \caption{\footnotesize TTSoC  overview}
%  \label{fig:a-ttsoc}
%\end{figure}

\end{document}